\documentclass[aps,twocolumn,showpacs]{revtex4}
\usepackage{dcolumn}
\usepackage{graphicx}
\usepackage{amsmath}
\usepackage{amsfonts}
\usepackage{amssymb}
\usepackage{psfrag}
\usepackage{wrapfig}
\usepackage{subfigure}
\usepackage{makeidx}
\usepackage{bm}
\usepackage{cases}

\begin{document}

\title{Rational W-shaped Optical Soliton on Continuous Wave in Presence of Kerr Dispersion and Stimulated Raman Scattering}
\author{Li-Chen Zhao$^{1}$}
\author{Sheng-Chang Li$^2$}\email{scli@mail.xjtu.edu.cn}
\author{Liming Ling$^{3}$}\email{lingliming@qq.com}

\address{$^1$Department of Physics, Northwest University, Xi'an
710069, China}
\address{$^2$School of Science, Xi'an Jiaotong University, Xi'an 710049, China}
\address{$^3$Department of Mathematics, South China University of Technology, Guangzhou 510640, China}

\date{July 29, 2013}
\begin{abstract}

We study localized wave on continuous wave background analytically in a nonlinear fiber with higher order effects such as higher order dispersion, Kerr dispersion,
and stimulated inelastic scattering. We present an exact rational W-shaped soliton solutions, whose structural properties depend on the frequency of the background field. The hump value increase with the decrease of the background frequency in the certain regime. The highest value of the W-shaped soliton can be nine times the background's, and the distribution shape is identical with the one of well-known eyes-shaped rogue wave with its maximum peak. The numerical stimulations indicate that the W-shaped soliton is stable with small perturbations.

\end{abstract}

\pacs{05.45.Yv, 42.65.Tg, 42.81.Dp}
 \maketitle

\section{Introduction}
Recently, Kuanetsov-Ma soliton, Akhmediev breather, and Peregrine soliton have been observed in nonlinear fiber \cite{Kibler,K-M,Dudley}, which provides a good platform to study dynamics of nonlinear localized waves on continuous wave (CW) background conveniently.
 These experimental studies show that the simplified  nonlinear Schr\"odinger(NLS) equation can describe the dynamics of localized waves well. For nonlinear fiber, the simplified NLS just contain
group velocity dispersion (GVD) and its counterpart self-phase
modulation (SPM). The propagation of femtosecond pulse is tempting and desirable to improve the capacity of high-bit-rate transmission systems. But for ultrashort pulses whose duration are shorter than $100fs$, in addition to the SPM,
the nonlinear susceptibility will produce higher order nonlinear
effects like the Kerr
dispersion (otherwise called the self-steepening) and the stimulated Raman scattering (SRS). Apart from
GVD, the ultrashort pulse will also suffer from third order
dispersion (TOD). These are the most general terms that have to be
taken into account when extending the applicability of the NLS
\cite{K. Porsezian,Kodama}. With these effects, the corresponding
integrable equation was derived as Sasa-Satsuma(S-S) equation
\cite{SS}.

 The studies on S-S model indicate that the
nonlinear waves in nonlinear fiber with these high-order effects are much more diverse than the
ones for simplified NLS model \cite{SS,Wright,Li,Mihalache,Mihalache2,Mihalache3}. Recently, it was found that the high-order effects could make the RW twisted, and the rational solutions of S-S equation had distinctive
properties from the ones of the well-known NLS equation \cite{rws,Chen}.  The Linearized Stability Analysis of S-S model in \cite{Wright} suggests that there are both modulational instability and stability regimes for low perturbation frequencies on the CW background. The rational solutions obtained in \cite{rws,Chen} are all in the modulational instability regime. Then, we can expect that the soliton solution could exist on the CW background in the modulational stability regime.

 In this
paper, we study on analytical rational solutions of the S-S equation through
Darboux transformation method. We present an exact
rational solution on CW background of S-S model, which corresponds to W-shaped soliton and involves in the modulational stability regime \cite{Wright}. It does not describe the dynamics of RW, in contrast to the rational ones of the simplified NLS, which have been used to describe RW phenomena photographically \cite{Ruban,N.Akhmediev,Kharif,Pelinovsky}. Its dynamics is similar with the W-shaped soliton reported in \cite{Li}. But the solution form is distinctive from the ones in \cite{Li} which has a nonrational form, and their distribution shapes are distinguished too. Furthermore, we discuss the stability of the rational W-shaped soliton through numerical stimulation method.

\section{The S-S model and continuous wave background}
According to the original work of Sasa and Satsuma \cite{SS},  the
evolution equations for the optical fields in a fiber with
the high-order effects mentioned above can be written as
\begin{eqnarray}
i E_{z} + \frac{1}{2} E_{tt}+ |E|^2 E +i \epsilon
[E_{ttt}+ 6 |E|^2 E_{t}+ 3 E
|E|^2_t]=0.
\end{eqnarray}
Here, an arbitrary real parameter $\epsilon $ scales the integrable
perturbations of the NLS equation. The units are dimension-less after performing proper scalar transformation. When $ \epsilon= 0$, Eq. (1) reduces to the standard NLS equation which has only
the terms describing lowest order dispersion and self-phase
modulation. The soliton solutions have been presented on the
 zero background in \cite{Gedalin,Gilson,Kim}. Here, we study rational solutions on a CW background,
\begin{eqnarray}\label{simptrans}
E_{0}(t,z)&=&c \exp{[i \theta_1]}\exp{[\frac{i}{6\epsilon}(t-\frac{z}{18\epsilon})]},
\end{eqnarray}
where
\begin{eqnarray}
\theta_1&=& w T+\epsilon w^3 z-6 \epsilon w
c^2 z,\nonumber
\end{eqnarray}
and $T=t-\frac{z}{12 \epsilon}$. $c$ denotes the
background amplitude. $w$ is the frequency of the optical background field.
Performing the Darboux transformation \cite{K. Nakkeeran} from the
above seed solution, one can derive kinds of localized waves
solution.  Notably, we find a new type rational solution on continuous background with some certain conditions on the background's amplitude and frequency $c\geq 2w$.
\begin{figure}[htb]
\centering
\includegraphics[width=80mm,height=65mm]{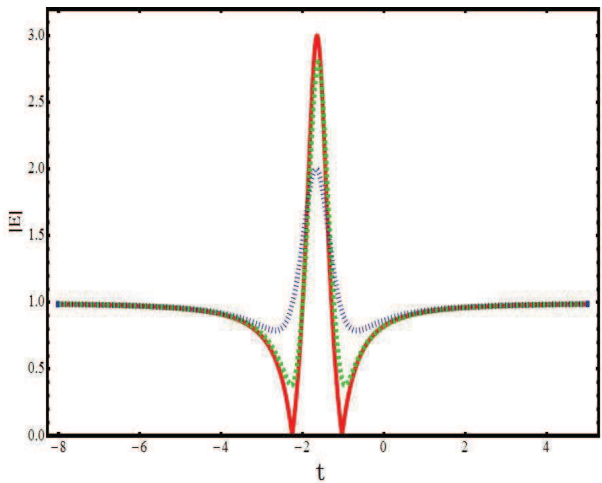}
\caption{(color online) The distribution shape of the rational solution $|E(t,z)|$ at $z=0$ with $w=0$ (red solid line), $w=0.3$ (green dashed line), and $w=0.5$ (blue dotted line). The parameters are $ c=1$ and  $\epsilon =0.1$. It is seen that the humps' value is inversely proportional to the value of the background frequency in the regime $0\leq w \leq \frac{c}{2}$. }
\end{figure}


\begin{figure}[htb]
\centering
\subfigure[]{\includegraphics[height=65mm,width=80mm]{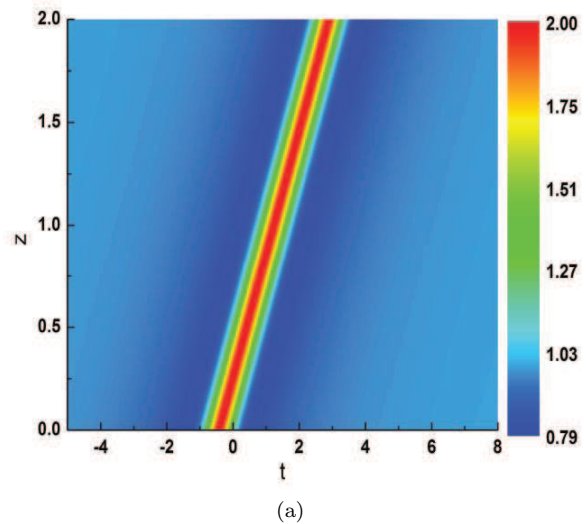}}
\hfil
\subfigure[]{\includegraphics[height=50mm,width=70mm]{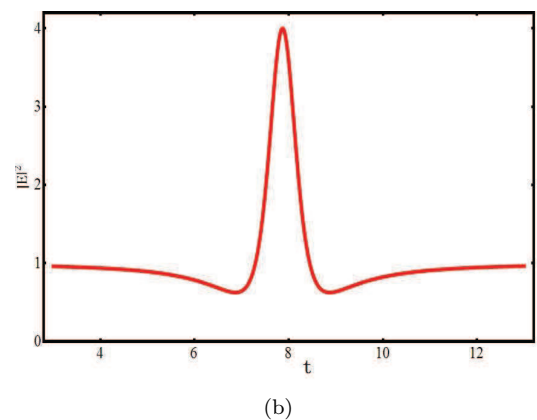}}
\caption{(color online) (a) The evolution of the rational solution $|E(t,z)|$ with $w=c/2$. (b) The cut plot of (a) at $z=5$. It is seen that the highest density value of $|E(t,z)|^2$ is four times the background's. The parameters are $ c=1$, $w=0.5$, and  $\epsilon =0.1$. }
\end{figure}
\begin{figure*}[htb]
\centering
\subfigure[]{\includegraphics[height=65mm,width=80mm]{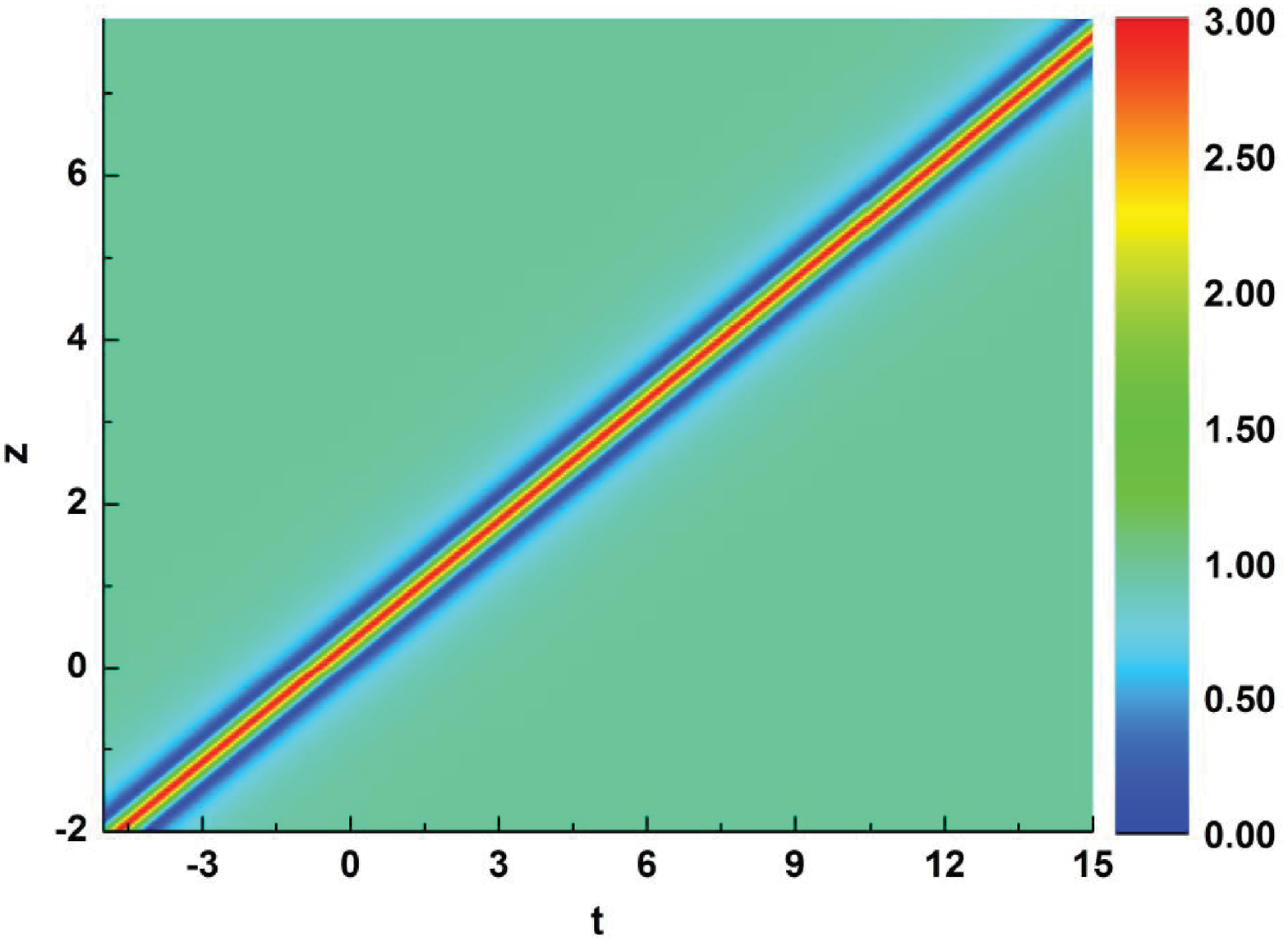}}
\hfil
\subfigure[]{\includegraphics[height=62mm,width=70mm]{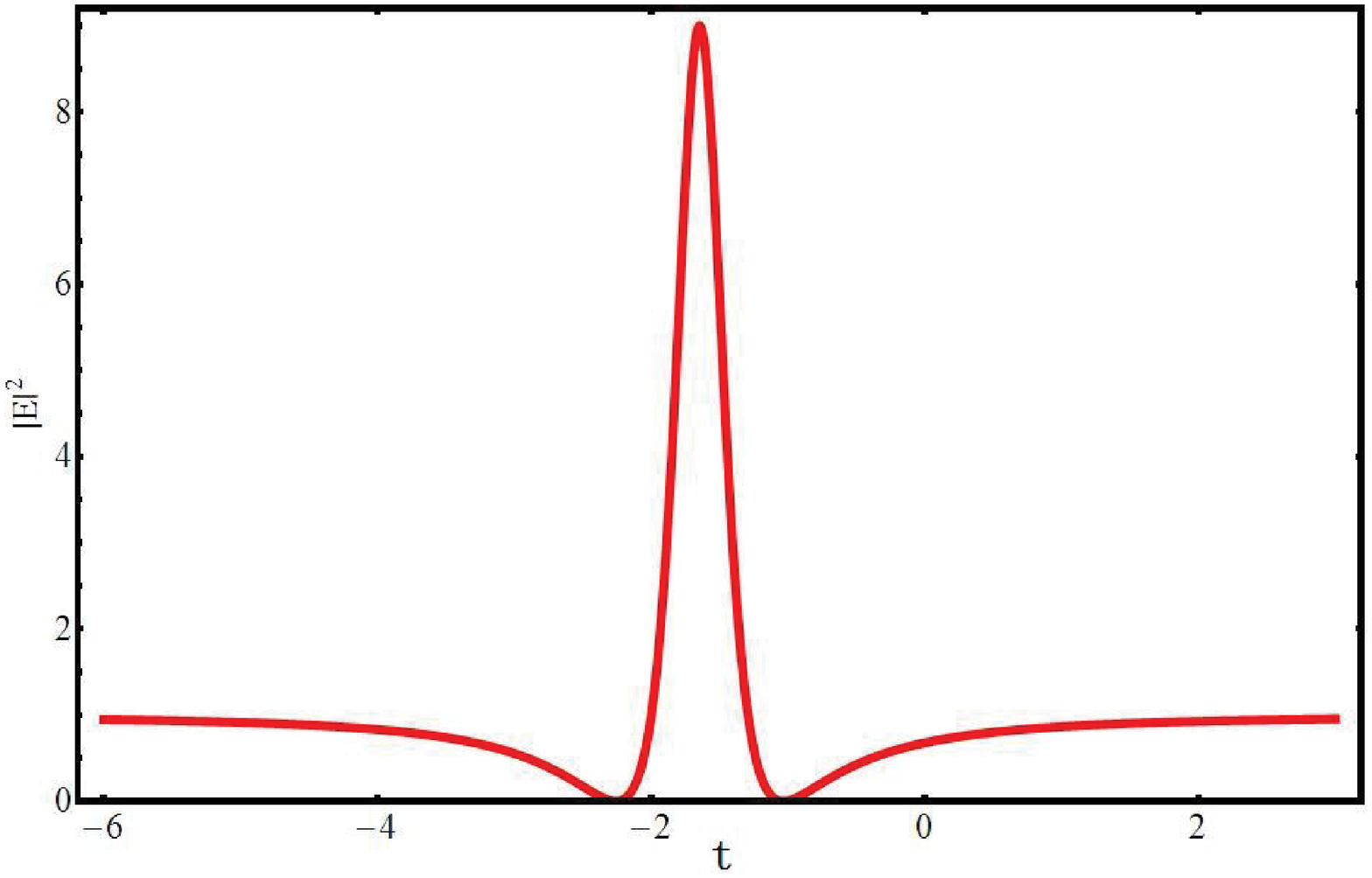}}
\caption{(color online) (a) The evolution of the rational solution $|E(t,z)|$ with $w=0$. (b)  The cut plot of (a) at $z=0$. It is seen that the highest density value of $|E(t,z)|^2$ is nine times the background's. The distribution structure is identical to the one of well-known eye-shaped rogue wave at its maximum peak emerging moment. The parameters are $ c=1$, $w=0$, and  $\epsilon =0.1$. }
\end{figure*}

We find that the rational solution does not correspond to rogue wave, in contrast to the ones of the simplified NLS \cite{Kibler,He,Yang,Ling}. Its dynamics corresponds to soliton's which has a stable distribution shape with evolution, and the distribution shape like a ``W" which has one hump and two valleys on the hump's two sides. Therefore, we call it rational W-shaped soliton. For S-S model, the instability regime are quite different from the simplified NLS, which just involves instability around low perturbation frequencies on continuous wave background. The Linearized Stability Analysis in \cite{Wright} suggests that there are both modulational instability and stability regimes for low perturbation frequencies on the continuous wave. Based on these results, we can qualitatively know that the rational solutions reported in \cite{rws,Chen} are in the modulational instability regime since the solution's dynamics corresponds to rogue wave behavior. Then, the rational solution obtained here should be involved with the modulational stability regime.
\begin{figure*}[htb]
\centering
\subfigure[]{\includegraphics[height=60mm,width=80mm]{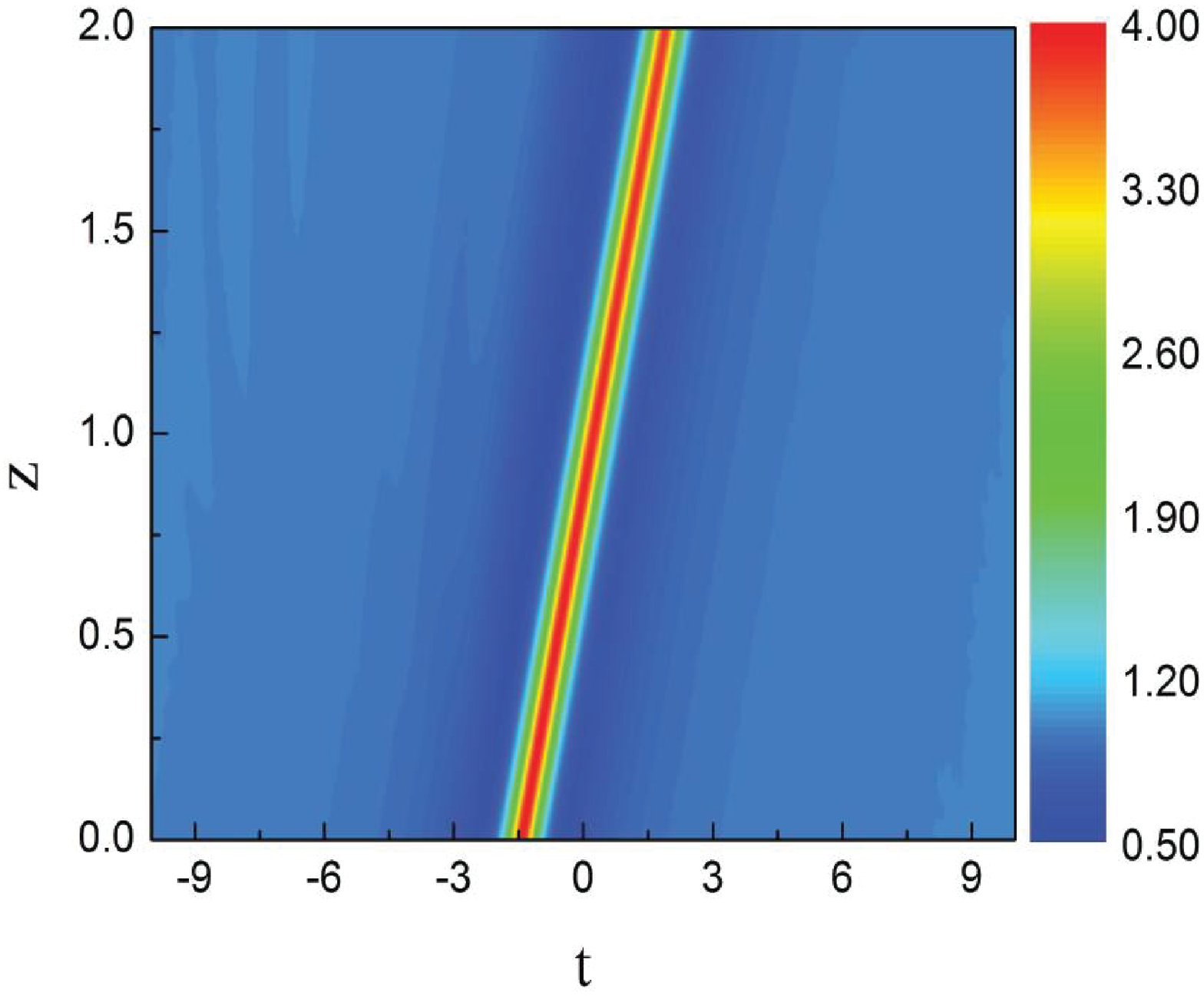}}
\hfil
\subfigure[]{\includegraphics[height=60mm,width=80mm]{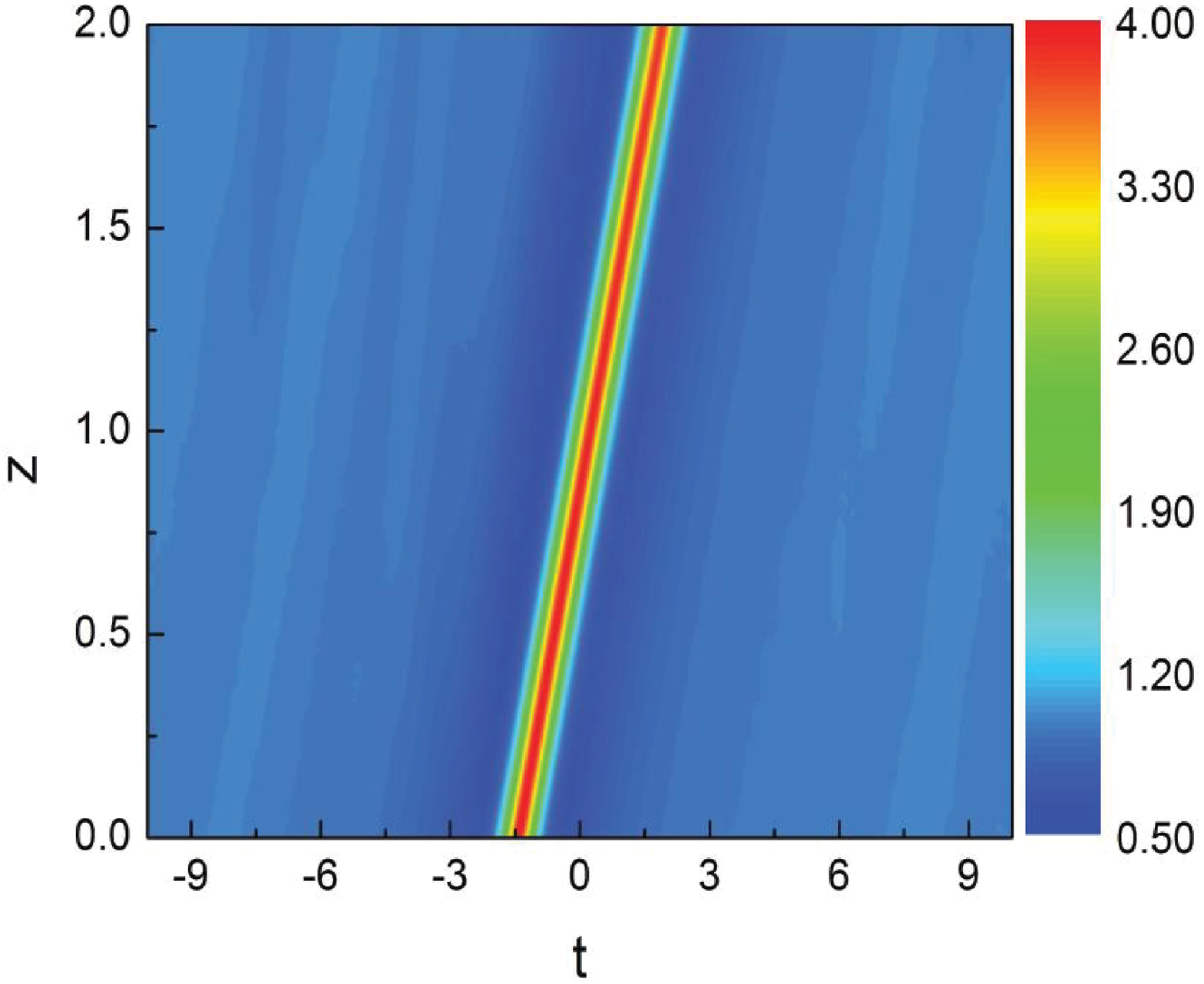}}
\caption{(color online) (a) The numerical evolution of $|E(t,z)|^2$ from the initial condition given by the rational solution $E(t,z=0)$ with $w=0$. The evolution agrees  well with the exact one in Fig. 2(a). (b) The numerical evolution of $|E(t,z)|^2$ from the initial condition $E(t,z=0)+0.02 sin(0.05 t)$. It is shown that the W-shaped soliton is robust with the small perturbations.  The parameters are $ c=1$, $w=0$, and  $\epsilon =0.1$. }
\end{figure*}

 Moreover, we find that the intensity values of the W-shaped soliton's hump and valleys depend on the value of the continuous wave background's frequency $w$. The hump's value increase with the decrease of $w$, for $0\leq w \leq \frac{c}{2}$, shown in Fig. 1. The lowest value of the hump can be four times the background's with $w=c/2$, and the biggest value of the hump can be nine times the background's with $w=0$. Since the generalized rational solution is quite complicated, we do not present it explicitly here.  However, we can obtain explicit and simpler form from the solution with some certain condition. As examples, we show two cases with $w=c/2$ and $w=0$ separately as follows.

\section{Two explicit cases for The rational W-shaped soliton solution}
Case 1: Under the condition $w=\frac{c}{2}$, the background amplitude can be set as $c=1$ without losing generality, the exact rational solution of Eq. (1) can be simplified as
\begin{widetext}
\begin{eqnarray}
E[t,z]&=&\left\{\frac{i \sqrt{3} -1}{2}+\frac{2 \left[-99 (\sqrt{3}+i) \epsilon z+12 (\sqrt{3}+i) T+(-3+\sqrt{3}+2\sqrt{3} i) (8-8 i)\right]}{-3267 \epsilon^2 z^2+8 T \left[99 \epsilon z+4 (\sqrt{3}-6)\right]-264 (\sqrt{3}-6) \epsilon z-48 T^2+32 (2 \sqrt{3}-7)} \right\}
\nonumber\\&&\cdot \exp{[\frac{i}{6 \epsilon}(t-\frac{z}{18\epsilon})+\frac{1}{8} i (4 T-23 z \epsilon )]}.
\end{eqnarray}
\end{widetext}
 The evolution of the rational solution is shown in Fig. 2. It is seen that there are two valleys and one hump which are kept very well with the propagation distance. The solution corresponds to a stable soliton solution although it has rational solution form. This is quite different from the rational solution presented in \cite{rws, Chen}. The soliton's shape is similar to the ``W"-shaped soliton presented in \cite{Li}. But the middle highest hump is much higher than the background, and its $|E|^2$ value is four times the background's density value, which is distinctive from the ones presented in \cite{Li}. To demonstrate these differences, we calculate the hump's value $|E|^2_h=4$ and its valleys' values are both $|E|^2_v=\frac{5}{8}$,  with $c=1$. Their corresponding trajectories are all straight lines on the temporal-spatial distribution plane. For the lines, $\frac{\partial t}{\partial z}=\frac{99 \epsilon ^2+1}{12 \epsilon }$, it is seen that the parameter $\epsilon$ determines the value, which suggests that these high-order effects affects the localized wave's velocity on the time.

Case 2: With $w=0$ and $c=1$, the generalized rational solution can be simplified
 as follows,
 \begin{widetext}
\begin{eqnarray}
E[t,z]&=&\left[-1+\frac{2}{4 T^2-2 T (48 z \epsilon +\sqrt{2}-4)+576 z^2 \epsilon ^2+24 (\sqrt{2}-4) z \epsilon -2 \sqrt{2}+5}\right]\cdot \exp{[\frac{i}{6 \epsilon}(t-\frac{z}{18\epsilon})]}.
\end{eqnarray}
\end{widetext}

Interestingly, the density peak of $|E|^2$ is nine times the backgrounds, and the minimum density value is nearly zero, shown in Fig. 3. The distribution shape is identical with the well-known fundamental RW solution with highest peak value of the simplified NLS \cite{Kibler,He,Yang,Ling}.  These characters are different from the fundamental W-shaped soliton in Case 1. Moreover, we find that the maximum value of the W-shaped soliton is nine times the background's with the condition $w=0$. The corresponding solutions with frequencies in the regime $0\leq w \leq \frac{c}{2}$  are simple rational solutions, which are all different from the results presented in \cite{rws,Chen}.

\section{discussion and conclusion}
 We present an exact rational solution of S-S equation, which can be used to describe W-shaped soliton in a nonlinear fiber with higher order effects such as higher order dispersion, Kerr dispersion,
and stimulated inelastic scattering.  Significantly, we find the hump value of the W-shaped soliton   depends on the frequency of the background field. The value is inversely proportional to the value of the background frequency in the regime $0\leq w \leq \frac{c}{2}$.  The highest value of the W-shaped soliton can be nine times the background's with $w=0$, and the distribution shape is identical with the one of well-known eyes-shaped rogue wave with its maximum peak. The lowest value of the soliton can be four times the background's with $w=c/2$, and the valleys' values are $5/8$ of the background's density value. Finally, we stimulate the rational W-shaped soliton through the split-step Fourier
method. The numerical results suggest that the soliton is stable under small perturbations. As an example, we show the case for the rational solution with $w=c/2$ and $c=1$ in Fig. 4. In Fig. 4(a), we stimulate the evolution of the initial pulse corresponding to the ones in Fig. 2(a) at $z=0$. It is seen that the stimulated one agree well with the exact one. Furthermore, we add some small perturbations on the initial pulse as $E(t,z=0)+ 0.02 sin(b t)$. The numerical results indicate that the W-shaped soliton is stable under the perturbations, such as the one in Fig. 4(b) with $b=0.05$.

\section*{Acknowledgments}
We are grateful to Prof. Zhan-Ying Yang for his helpful discussions.
This work is supported by the Natural Science Foundation
of China (Grant Nos. 11305120 and 11005055), the National Fundamental Research Program of
China (Contact 2011CB921503, 2013CB834100)

\end{document}